\documentclass[draftcls, onecolumn]{IEEEtran}
\usepackage{pdfsync,srcltx}
\usepackage{amsmath}
\usepackage{amsfonts}
\usepackage{amssymb}
\usepackage{graphicx,subfigure,psfrag}
\usepackage{url,ifthen}
\usepackage{pmat}
\usepackage{cite}
\usepackage[ieee]{jphmacros2e}

\usepackage[norefs,nocites]{refcheck}
\makeatletter\let\NAT@parse\undefined\makeatother 

\newcommand{\R}{\mathbb{R}}

\newcommand{\N}{\mathbb{N}}
\newcommand{\Z}{\mathbf{Z}}
\renewcommand{\L}{\mathcal{L}}

\newcommand{\scr}{\mathcal}
\newcommand{\comment}[1]{}

\allowdisplaybreaks

\title{{\LARGE 
Control of large 1D networks of double integrator agents: role of heterogeneity and asymmetry on stability margin}}

\author{He Hao and Prabir Barooah
\thanks{He Hao and Prabir Barooah are with Department of Mechanical and Aerospace Engineering,
        University of Florida, Gainesville, FL 32611, USA
        {\tt\small {hehao,pbarooah}@ufl.edu}. This work was supported by the National Science Foundation through Grants CNS-0931885 and ECCS-0925534, and by the Institute for Collaborative Biotechnologies through grant DAAD19-03-D-0004.}
	\thanks{	The conference version of this paper appeared in~\cite{HH_PB_CDC:10}.}
}

\begin{document}
\maketitle
\thispagestyle{empty}
\pagestyle{empty}

\begin{abstract}
We consider the distributed control of a network of heterogeneous agents with double integrator dynamics to maintain a rigid formation in 1D Euclidean space. The control signal at each vehicle is allowed to use relative position and velocity with its two nearest neighbors. Most of the work on this problem, though extensive, has been limited to homogeneous networks, in which agents have identical mass and controller, and symmetric control, in which information from front and back neighbors are weighted equally. We examine the effect of heterogeneity  and asymmetry on the closed loop stability margin, which is measured by the real part of the least stable pole of the closed-loop system. By using a PDE (partial differential equation) approximation in the limit of large number of vehicles, we show that heterogeneity has little effect while asymmetry has a significant effect on the stability margin. When control is symmetric, the stability margin decays to $0$ as $O(1/N^2)$, where $N$ is the number of agents, even when the agents are heterogeneous in their masses and control gains. In contrast, we show that arbitrarily small amount of asymmetry in the velocity feedback gains can improve the decay of the stability margin to $O(1/N)$. Poor design of such asymmetry makes the closed loop unstable for sufficiently large $N$. Moreover, if there is equal amount of asymmetry in both position and velocity feedback gains, the stability margin of the network can be bounded away from $0$, uniformly in $N$. This results thus eliminates the degradation of closed-loop stability margin with increasing $N$, but its sensitivity to external disturbances becomes much worse than symmetric control. Numerical computations  are provided to corroborate the analysis.  
\end{abstract}

\section{Introduction}\label{sec:intro}
In this paper we examine the closed loop dynamics of a network consisting of $N$ interacting agents arranged in a line, where the agents are modeled as double integrators and each agent interacts with its two nearest neighbors (one on either side) through its local control action. This is a problem that is of primary interest to formation control applications, especially to platoons of vehicles, where the vehicles are modeled as point masses. An extensive literature exists on 1-D automated platoons; see~\cite{chu_platoon,darbha1994comparison,zhang1999using,DS_KH_CC_PI_VSD:94,Seiler_disturb_vehicle_TAC:04,PB_PM_JH_TAC:09} and references therein. In the vehicular platoon problem, the formation try to track a  desired trajectory while maintaining a rigid formation geometry. The desired trajectory of the entire vehicular platoon is given in terms of trajectory of a fictitious reference vehicle, and the desired formation geometry is specified in terms of constant inter-vehicle spacings.

Although significant amount of research has been conducted on robustness-to-disturbance and stability issues of double integrator networks with decentralized control, most investigations consider the homogeneous case in which each agent has the same mass and employs the same controller (exceptions include~\cite{KhatirDavison_NonIdenticalK_ACC:04,HT_DC_RAS:07,RM_JB_TAC:09}). In addition, only symmetric control laws are considered in which the information from both the neighboring agents are weighted equally, with~\cite{veerman_stability,PB_PM_JH_TAC:09} being exceptions. Khatir \textit{et. al.} proposes heterogeneous control gains to improve string stability (sensitivity to disturbance) at the expense of control gains increasing without bound as $N$ increases~\cite{KhatirDavison_NonIdenticalK_ACC:04}. Middleton~\textit{et. al.} considers both unidirectional and bidirectional control, and concludes heterogeneity has little effect on the string stability under certain conditions on the high frequency behavior and integral absolute error~\cite{RM_JB_TAC:09}. On the other hand,~\cite{veerman_stability} examines the effect of asymmetry (but not heterogeneity) on the response of the platoon as a result of sinusoidal disturbances in the lead vehicle, and concludes the asymmetry makes sensitivity to such disturbances worse. 

In this paper we analyze the case when the agents are \emph{heterogeneous} in their masses and control laws used, and also allow asymmetry in the use of front and back information. A decentralized \emph{bidirectional} control law is considered that uses only relative position and relative velocity information from the nearest neighbors. We examine the effect of heterogeneity and asymmetry on the stability margin of the closed loop, which is measured by the absolute value of the real part of the least stable pole. The stability margin determines the decay rate of initial formation keeping errors. Such errors arise from poor initial arrangement of the agents. The main result of the paper is that in a decentralized bidirectional control strategy, heterogeneity has little effect on the stability margin of the overall closed loop, while even small asymmetry can have a significant impact. In particular, we show that in the symmetric case, the stability margin decays to $0$ as $O(1/N^2)$, where $N$ is the number of agents. We also show that the asymptotic scaling trend of stability margin is not changed by agent-to-agent heterogeneity as long as the control gains do not have front-back asymmetry. On the other hand, arbitrary small amount of asymmetry in the way the local controllers use front and back information can improve the stability margin by a considerable amount. When each agent  weighs the relative velocity information from its front neighbor more heavily than the one behind it, the stability margin scaling trend can be improved from $O(1/N^2)$ to $O(1/N)$. In contrast, if more weight is given to the relative velocity information with the neighbor behind it, the closed loop becomes unstable if $N$ is sufficiently large. In addition, when there is equal amount of asymmetry in position and velocity feedback gains, the closed-loop is exponential stable for arbitrary finite $N$, and the stability margin can be uniformly bounded with the size of the network. This result makes it possible to design the control gains so that the stability margin of the system satisfies a pre-specified value irrespective of how many vehicles are in the formation.  However, in this special case, the sensitivity to disturbance becomes much worse than symmetric control. In contrast, with judicious asymmetry in velocity feedback alone improves the sensitivity to external disturbance. 

In this paper, we propose a PDE approximation to the coupled system of ODEs that model the closed loop dynamics of the network. This is inspired by the work~\cite{PB_PM_JH_TAC:09} that examined stability margin of 1-D vehicular platoons in a similar framework. Compared to~\cite{PB_PM_JH_TAC:09}, this paper makes two novel contributions. First, we consider heterogeneous agents (the mass and control gains vary from agent to agent), whereas~\cite{PB_PM_JH_TAC:09} consider only homogeneous agents. Secondly,~\cite{PB_PM_JH_TAC:09} considered the scenario in which the desired trajectory of  the platoon was one with a constant velocity, and moreover, every agent knew this desired velocity. In contrast, the control law we consider requires agents to know only the desired inter-agent separation; the overall trajectory information is made available only to agent $1$. This makes the model more applicable to practical formation control applications. It was shown in~\cite{PB_PM_JH_TAC:09} for the homogeneous formation that asymmetry in the position feedback can improve the stability margin from $O(1/N^2)$ to $O(1/N)$ while the absolute velocity feedback gain did not affect the asymptotic trend. In contrast, we show in this paper that with relative position and relative velocity feedback, asymmetry in the velocity feedback gain alone and in both position and velocity feedback gains are very important. The stability margin can be improved considerably by a judicious choice of asymmetry. 

Although the PDE approximation is valid only in the limit $N \to \infty$, numerical comparisons with the original state-space model shows that the PDE model provides accurate results even for small $N$ ($5$ to $10$). PDE approximation is quite common in many-particle systems analysis in statistical physics and traffic-dynamics (see the article~\cite{Helbing-review:01} for an extensive review.). The usefulness of PDE approximation in analyzing multi-agent coordination problems has been recognized also by researchers the controls community; see~\cite{AS_RS_CDC:09,EJ_PK_CDC:03,PB_PM_JH_TAC:09,HH_PB_PM_DSCC:09} for examples. A similar but distinct framework based on partial \emph{difference} equations has been developed by Ferrari-Trecate~\textit{et. al.}~\cite{GFT_AB_MG_TAC:06}.

The rest of this paper is organized as follows. Section~\ref{sec:results} presents the problem statement and the main results of this paper. Section~\ref{sec:problem} describes the PDE model of the network of agents. Analysis and control design results together with their numerical corroboration appear in Sections~\ref{sec:instability-analysis} and~\ref{sec:mistuning}, respectively.  The paper ends with a summary in Section~\ref{sec:conc}.

\section{Problem statement and main results}\label{sec:results}

\subsection{Problem statement}\label{sec:statement}
\begin{figure}
	  \psfrag{O}{$O$}
	  \psfrag{o}{$0$}
	  \psfrag{l}{$1$}
	  \psfrag{X}{$X$}
	  \psfrag{x}{$x$}
	  \psfrag{d1}{\scriptsize$\Delta_{0,1}$}
	  \psfrag{d2}{\scriptsize$\Delta_{N-1,N}$}
	   \psfrag{v}{$v^{*}\;t$}
	   \psfrag{d}{\scriptsize $1/N$}
	   \psfrag{0}{\scriptsize $0$}
	   \psfrag{1}{\scriptsize $1$}
	   \psfrag{N-1}{\scriptsize $N-1$}
	   \psfrag{N}{\scriptsize $N$}
	   \psfrag{Di}{\scriptsize Dirichlet}
	   \psfrag{Ne}{\scriptsize Neumann}
\centering
 \subfigure[A pictorial representation of 1D network.]{\includegraphics[scale = 0.27]{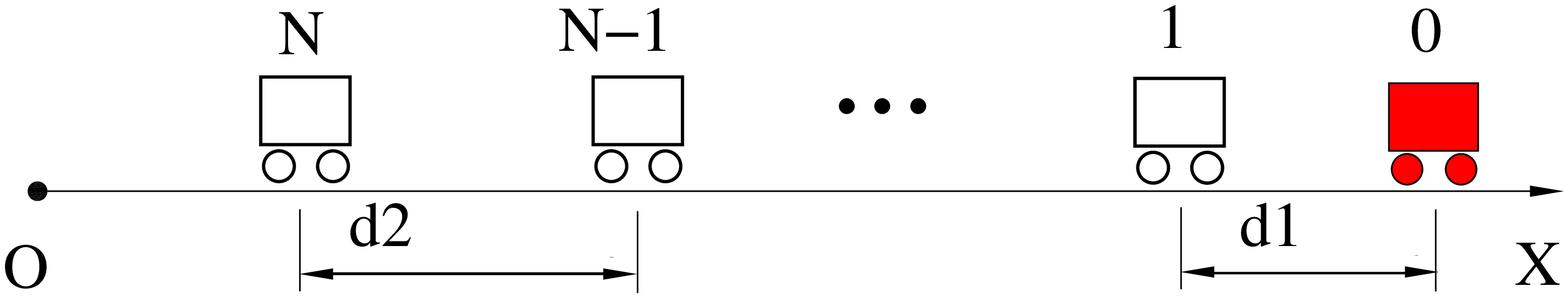}} \quad 
 \subfigure[A Redrawn graph of the same network.]{\includegraphics[scale = 0.27]{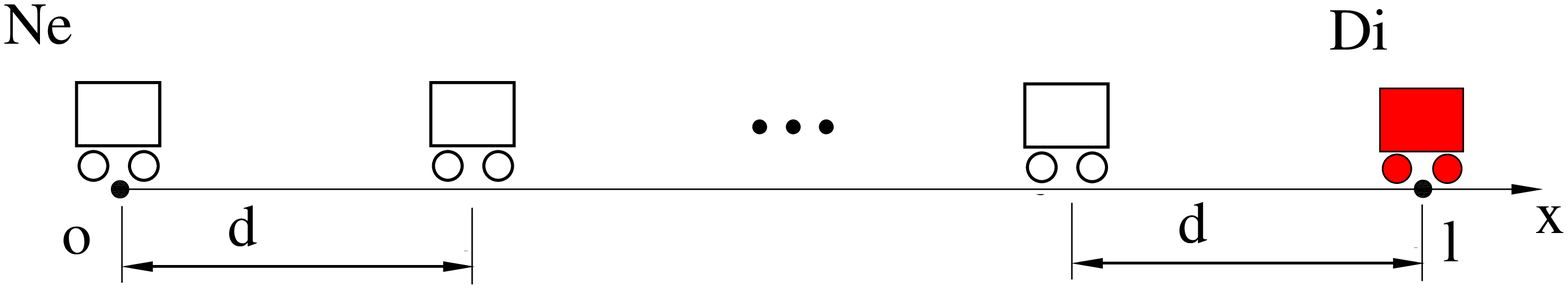}} 
\caption{Desired geometry of a network with $N$ agents and $1$ "reference agent", which are moving in 1D Euclidean space. The filled agent in the front of the network represents the reference agent, it is denoted by "$0$". (a) is the original graph of the network in the $p$ coordinate and (b) is the redrawn graph of the same network in the $\tilde{p}$ coordinate. }
\label{fig:fig1}
\end{figure}

We consider the formation control of $N$ heterogeneous agents which are moving in 1D Euclidean space, as shown in Figure~\ref{fig:fig1} (a). The position and mass of each agent are denoted by  $p_i$ and $m_i$ respectively. The mass of each agent is bounded, $|m_i-m_0| /m_0 \leq \delta$ for all $i$, where $m_0>0$ and $\delta \in [0,1)$ are constants. The dynamics of each agent are modeled as a double integrator:
\begin{align}\label{eq:vehicle-dynamics}
 m_i \ddot{p}_i= u_i,
 \end{align}
where $u_i$ is the control input (acceleration or deceleration command).  This is a commonly used model for vehicle dynamics in studying vehicular formations, which results from feedback linearization of actual non-linear vehicle dynamics~\cite{darbha1994comparison,feedback_linearization}.

The desired trajectory of the formation is given in terms of  a \emph{fictitious} reference agent with index $0$ whose trajectory is denoted by $p^*_0(t)$. Since we are interested in translational maneuvers of the formation, we assume the desired trajectory is a constant-velocity type, i.e. $p^*_0(t)=v_0t+c_0$ for some constants $v_0$ and $c_0$. The information on the desired trajectory of the network is provided only to agent $1$. The desired geometry of the formation is specified by the \emph{desired gaps} $\Delta_{i-1,i}$ for $i=1,\dots,N$, where $\Delta_{i-1,i}$ is the desired value of $p_{i-1}(t) - p_{i}(t)$. The control objective is to maintain a rigid formation, i.e., to make neighboring agents maintain their pre-specified desired gaps and to make agent $1$ follow its desired trajectory $p^*_0(t) - \Delta_{0,1}$. Since we are only interested in maintaining rigid formations that do not change shape over time, $\Delta_{i-1,i}$'s are positive constants.

In this paper, we consider the following \emph{decentralized} control law, whereby the control action at the $i$-th agent depends on i) the {\em relative position measurements}  ii) the {\em relative velocity measurements} with its immediate neighbors in the formation:
\begin{align}\label{eq:control-lawd}
	u_{i} = &-k_{i}^{f}(p_{i}-p_{i-1}+\Delta_{i-1,i})-k_{i}^{b}(p_{i}-p_{i+1}-\Delta_{i,i+1}) -b_{i}^{f}(\dot{p}_{i}-\dot{p}_{i-1})-b_{i}^{b}(\dot{p}_{i}-\dot{p}_{i+1}),
\end{align}
where $i=\{1,\dots,N-1 \}$, $k_{(.)}^{f}, k_{(.)}^{b}$ are the front and back position gains and $b_{(.)}^{f}, b_{(.)}^{b}$ are the front and back velocity gains respectively. For the agent with index $N$ which does not have an agent behind it, the control law is slightly different:
\begin{align}\label{eq:control-law-N}
	u_{N} = &-k_{N}^{f}(p_{N}-p_{N-1}+\Delta_{N-1,N})-b_{N}^{f}(\dot{p}_{N}-\dot{p}_{N-1}).
\end{align}
Each agent $i$ knows the desired gaps $\Delta_{i-1,i}$ and $\Delta_{i,i+1}$, while only agent $1$ knows the desired trajectory $p_0^*(t)$ of the fictitious reference agent.

Combining the open loop dynamics~\eqref{eq:vehicle-dynamics} with the control law~\eqref{eq:control-lawd}, we get
\begin{align}\label{eq:state_blah_blah_1}
	m_i \ddot{p}_i  = &-k_{i}^{f}(p_{i}-p_{i-1}+\Delta_{i-1,i})
	-k_{i}^{b}(p_{i}-p_{i+1}-\Delta_{i,i+1}) -b_{i}^{f}(\dot{p}_{i}-\dot{p}_{i-1})-b_{i}^{b}(\dot{p}_{i}-\dot{p}_{i+1}), \end{align}
where $i\in \{1,\dots,N-1\}.$ The dynamics of the $N$-th agent are obtained by combining~\eqref{eq:vehicle-dynamics} and~\eqref{eq:control-law-N}, which are slightly different from~\eqref{eq:state_blah_blah_1}. The desired trajectory of the $i$-th agent is $  p_i^*(t) \defeq p_0^*(t) -  \Delta_{0,i}=p_0^*(t) - \sum_{j=1}^{i} \Delta_{j-1,j} $.  To facilitate analysis, we define the following tracking error:
\begin{align}\label{eq:state_error}
	\tilde{p}_i  & \eqdef p_i-p_i^*  & \Rightarrow & & \dot{\tilde p}_i& =\dot{p}_i-\dot{p}_i^*.    
\end{align}
Substituting~\eqref{eq:state_error} into~\eqref{eq:state_blah_blah_1}, and using $p_{i-1}^*(t) - p_{i}^*(t) = \Delta_{i-1,i}$, we get
\begin{align}\label{eq:new_dynamics}
	m_i \ddot{\tilde{p}}_{i}  = -k_{i}^f(\tilde{p}_{i}-\tilde{p}_{i-1})-k_{i}^b(\tilde{p}_{i}-\tilde{p}_{i+1})-b_{i}^f(\dot{\tilde{p}}_{i}-\dot{\tilde{p}}_{i-1})-b_{i}^b(\dot{\tilde{p}}_{i}-\dot{\tilde{p}}_{i+1}) .
\end{align}
 By defining the state $\psi \eqdef[\tilde{p}_{1},\dot{\tilde p}_{1}, \tilde{p}_{2},\dot{\tilde p}_{2},\cdots,\tilde{p}_{N}, \dot{\tilde p}_{N}]^{T}$,  the closed loop dynamics of the network can now be written
compactly from~\eqref{eq:new_dynamics} as:
\begin{align}\label{eq:closedloop-wholeplatoon}
\dot{\psi}=\mbf{A} \psi
\end{align}
where $\mbf A$ is the closed-loop state matrix and we have used the fact that $\tilde{p}_{0}(t) = \dot{\tilde{p}}_{0}(t) \equiv 0$ since the trajectory of the reference agent is equal to its desired trajectory.
\subsection{Main results}\label{sec:result-nominal}
The first two results rely on the analysis of the following PDE (partial differential equation) model of the network, which is seen as a continuum approximation of the closed-loop dynamics~\eqref{eq:new_dynamics}. The details of derivation of the PDE model are given in Section~\ref{sec:problem}. The PDE is given by
\begin{align}\label{eq:PDE_pre}
	m(x) \frac{\partial^2  \tilde{p}(x,t)}{\partial t^2}   = \Big (\frac{k^{f-b}(x)}{N}\frac{\partial}{\partial x}+\frac{k^{f+b}(x)}{2{N}^2}\frac{\partial ^2}{\partial {x}^2}+ \frac{b^{f-b}(x)}{N}\frac{\partial^2}{\partial x \partial t}+\frac{b^{f+b}(x)}{2{N}^2}\frac{\partial ^3}{\partial {x}^2 \partial t} \Big ) \tilde{p}(x,t),
\end{align}
with boundary condition: 
\begin{align}\label{eq:BC_pre}
	&\tilde{p}(1,t) = 0, & \frac{\partial \tilde{p}}{\partial x}(0,t) = 0,
\end{align}
where  $k^{f-b}(x), k^{f+b}(x), b^{f-b}(x)$ and $b^{f+b}(x)$ are defined as follows:
\begin{align*}  k^{f+b}(x) & \eqdef k^{f}(x) + k^{b}(x),  &  k^{f-b}(x) & \eqdef k^{f}(x) - k^{b}(x),\notag \\
  b^{f+b}(x) & \eqdef b^{f}(x)+ b^{b}(x),  &  b^{f-b}(x) & \eqdef b^{f}(x)- b^{b}(x),
\end{align*}
and $m(x), k^f(x),  k^b(x), b^f(x), b^b(x)$ are respectively the continuum approximation of $m_i, k^f_i,k^b_i,b^f_i,b^b_i$ of each agent with the following stipulation:
\begin{align}\label{eq:gains}
k_{i}^{ f \text{ or }b} &=k^{ f \text{ or }b} (x)|_{x =\frac{N-i}{N}}, & b_{i}^{ f \text{ or }b} &=b^{ f \text{ or }b} (x)|_{x = \frac{N-i}{N}}, & m_i=m(x)|_{x = \frac{N-i}{N}}.
\end{align}

We formally define symmetric control and stability margin before stating the first main result, i.e. the role of heterogeneity on the stability margin of the network.
\begin{definition}
The control law~\eqref{eq:control-lawd} is \emph{symmetric} if each agent uses the same front and back control gains: $k_{i}^{f}=k_{i}^{b}$ and $b_{i}^{f}=b_{i}^{b}$, for all $i \in \{1, 2,\dots, N-1 \}$. \frqed
\end{definition}
\begin{definition}
The stability margin of a closed-loop system, which is denoted by $S$, is the absolute value of the real part of the least stable pole. \frqed
\end{definition}

\begin{theorem}\label{thm:symmetric}
	Consider the PDE model~\eqref{eq:PDE_pre} of the network with boundary condition~\eqref{eq:BC_pre}, where the mass and the control gain profiles satisfy $|m(x)-m_0|/m_0 \leq \delta $, $|k^{(\cdot)}(x)-k_0|/k_0 \leq \delta $ and $|b^{(\cdot)}(x)-b_0|/b_0 \leq \delta $ for all $x \in [0,1]$ where $m_0, k_0$ and $b_0$ are positive constants, and $\delta \in [0,1)$ denotes the percent of heterogeneity. With symmetric control, the stability margin $S$ of the network satisfies the following:
\begin{align}\label{eq:less_stable_eigenvalue}
	(1-2\delta)\frac{\pi^2 b_0}{8m_0}\frac{1}{N^{2}} \leq S \leq (1+2\delta) \frac{\pi^2 b_0}{8m_0}\frac{1}{N^{2}},
\end{align}
when $\delta \ll 1$.\frqed
\end{theorem}

The result above is also provable for an arbitrary $\delta<1$ (not necessarily small) when the position gain is proportional to the velocity gain using standard results of Sturm-Liouville theory~\cite[ Chapter 5]{haberman}. 
For that case, the result is given in the following lemma and its proof is given in the end of the Appendix. 

\begin{lemma}\label{lem:hetero_mass}
Consider the PDE model~\eqref{eq:PDE_pre} of the network with boundary condition~\eqref{eq:BC_pre}. Let the mass and the control gains satisfy $0 < m_\mrm{min} \leq m(x) \leq m_\mrm{max}$, $0 < b_\mrm{min} \leq b^{f}(x)=b^{b}(x)=b(x) \leq b_\mrm{max}$ and $k^{f}(x)=k^{b}(x)=k(x)=\rho b(x)$ for all $x \in [0,1]$, where $m_\mrm{min}, m_\mrm{max}, b_\mrm{min}, b_\mrm{max}$ and $\rho$ are positive constants. The stability margin $S$ of the network satisfies the following:
\begin{align*}
	\frac{\pi^2 b_\mrm{min}}{8m_\mrm{max}}\frac{1}{N^{2}} \leq S \leq \frac{\pi^2 b_\mrm{max}}{8m_\mrm{min}}\frac{1}{N^{2}}. \tag*{\frqed}
\end{align*} 
\end{lemma}

\medskip

The main implication of the result above is that \emph{heterogeneity of masses and control gains plays no role in the asymptotic trend of the stability margin with $N$ as long as the control gains are symmetric}. Note that the $O(1/N^2)$ decay of the stability margin described above has been shown for homogeneous platoons (all agents have the same mass and use the same control gains) independently in~\cite{veerman_automated}, although the dynamics of the last vehicle are slightly different from ours. A similar result for homogeneous platoons with relative position and absolute velocity feedback was also established in~\cite{PB_PM_JH_TAC:09}. 

The second main result of this work is that the stability margin can be greatly improved by introducing front-back asymmetry in the \emph{velocity}-feedback gains. We call the resulting design \emph{mistuning}-based design because it relies on small changes from the nominal symmetric gain $b_0$. In addition, a poor choice of such asymmetry can also make the closed loop unstable.  Since heterogeneity is seen to have little effect, and for ease of analysis, we let $m_i=m_0$ in the sequel.

\begin{theorem}\label{thm:mistuning}
For an $N$-agent network with PDE model~\eqref{eq:PDE_pre} and boundary condition~\eqref{eq:BC_pre}. Let $m(x) = m_0$ for all $x \in [0,1]$, consider the problem of maximizing the stability margin by choosing the control gains with the constraint $| b^{(.)}(x) - b_0 |/b_0 \leq  \varepsilon$, where $\varepsilon$ is a positive constant, and $k^{(f)}(x) = k^{(b)}(x) = k_0$. If $\varepsilon \ll 1$, the optimal velocity gains are  
\begin{align}\label{eq:optimal-mistuned-gains}
	b^{f}(x)&=(1+\varepsilon)b_0, & b^{b}(x)&=(1-\varepsilon)b_0,
\end{align}
which result in the stability margin
\begin{align}\label{eq:stability-margin-mistuned}
	S =\frac{\varepsilon b_0}{m_0}\frac{1}{N} + O(\frac{1}{N^2})=O(\frac{1}{N}).
\end{align}
The formula is asymptotic in the sense that it holds for large $N$ and small $\varepsilon$. In contrast, for the following choice of asymmetry
\begin{align}\label{eq:bad-mistuned-gains}
	b^{f}(x)=(1-\varepsilon)b_0\quad  \quad b^{b}(x)&=(1+\varepsilon)b_0,
\end{align}
where $\varepsilon \ll 1$ is an small positive constant, the closed loop becomes unstable for sufficiently large $N$. \frqed
\end{theorem}

The theorem says that with arbitrary small change in the front-back asymmetry, so that velocity information from the front is weighted more heavily than the one from the back, the stability margin can be improved significantly over symmetric control. On the other hand, if velocity information from the back is weighted more heavily than that from the front,  the closed loop will become unstable if the network is  large enough. It is interesting to note that the optimal gains turn out to be homogeneous, which again indicates that heterogeneity has little effect on the stability margin. 

The astute reader may inquire at this point what are the effects of introducing asymmetry in the position-feedback gains while keeping velocity gains symmetric, or introducing asymmetry in both position and velocity feedback gains. It turns out when equal asymmetry in both position and velocity feedback gains are introduced, the closed loop is exponentially stable for arbitrary $N$. Moreover, the stability margin scaling trend can be uniformly bounded below in $N$ when more weights are given to the information from its front neighbor. We state the result in the next theorem. 

\begin{theorem}\label{thm:equal-asymmetry}
	For an $N$-agent network with PDE model~\eqref{eq:PDE_pre} and boundary condition~\eqref{eq:BC_pre}. Let $m(x) = m_0$ for all $x \in [0,1]$. With the following asymmetry in control
$k^f(x)=(1+\varepsilon)k_0$, $k^b(x)=(1-\varepsilon)k_0$, $b^f(x)=(1+\varepsilon)b_0$, $b^b(x)=(1-\varepsilon)b_0$, where $\varepsilon$ is the amount of asymmetry satisfying $\varepsilon \in (0,1)$,  the stability margin of the network can be uniformly bounded below as follows:
\begin{align*}
	S\geq	\min \Big \{ \frac{b_0 \varepsilon^2 }{2}, \frac{k_0}{b_0} \Big\} = O(1). \tag*{\frqed}
\end{align*}
\end{theorem}

\smallskip

This asymmetric design therefore makes the resulting control law highly scalable; it eliminates the degradation of closed-loop stability margin with increasing $N$.  It is now possible to design the control gains so that the stability margin of the system satisfies a pre-specified value irrespective of how many vehicles are in the formation. The result above is for equal amount  of asymmetry in the position feedback and velocity feedback gains. This constraint of equal asymmetry in position and velocity feedback is imposed in order to make the analysis tractable. The analysis of the stability margin in the following cases are open problems: (i) unequal asymmetry in position and velocity feedback, (ii) velocity feedback gains are kept at their nominal symmetric values and asymmetry is introduced in the position feedback gains only.

\section{PDE model of the closed-loop dynamics}\label{sec:problem}
In this paper, all the analysis and design is performed using a PDE model, whose results are validated by numerical computations using the state-space model~\eqref{eq:closedloop-wholeplatoon}. We now derive a continuum approximation of the coupled-ODEs~\eqref{eq:new_dynamics} in the limit of large $N$, by following the steps involved in a finite-difference discretization in reverse.  We define
 \begin{align*}
 	&k_i^{f+b} \eqdef  k_{i}^f+ k_{i}^b,  &k_i^{f-b} \eqdef  k_{i}^f - k_{i}^b, \notag \\ & b_i^{f+b} \eqdef  b_{i}^f+b_{i}^f, &b_i^{f-b} \eqdef  b_{i}^f- b_{i}^b.
 \end{align*}
 Substituting these into~\eqref{eq:new_dynamics}, we have
\begin{align}\label{eq:prePDExx}
	m_i \ddot{\tilde{p}}_{i} =& 
      -\frac{k_i^{f+b}+ k_i^{f-b}}{2}(\tilde{p}_{i}-\tilde{p}_{i-1}) -\frac{k_i^{f+b}- k_i^{f-b}}{2}(\tilde{p}_{i}-\tilde{p}_{i+1}) \notag \\ &-\frac{b_i^{f+b}+ b_i^{f-b}}{2}(\dot{\tilde{p}}_{i}-\dot{\tilde{p}}_{i-1}) - \frac{b_i^{f+b}- b_i^{f-b}}{2}(\dot{\tilde{p}}_{i}-\dot{\tilde{p}}_{i+1}).
\end{align}
To facilitate analysis, we redraw the graph of the 1D network, so that each vehicle in the new graph is drawn in the interval $[0,1]$, irrespective of the number of agents. The $i$-th agent in the ``original'' graph, is  now drawn at position $(N-i)/N$ in the new graph. Figure~\ref{fig:fig1} shows an example.

The starting point for the PDE  derivation is to consider a function
$\tilde{p}(x,t): [0 , 1] \times [0, \; \infty ) \to \R$ that satisfies:
\begin{align}\label{eq:p_approx}
  \tilde{p}_{i}(t) = \tilde{p}(x,t)|_{x = (N-i)/N}, 
\end{align}
such that functions that are defined at discrete points $i$ will be approximated by functions that are defined everywhere in $[0,1]$. The original functions are thought of as samples of their continuous approximations. We formally introduce the following scalar functions  $k^f(x),k^b(x),b^f(x),b^b(x)$ and $m(x):[0,1]\to \R$  defined according to the stipulation:
\begin{align}\label{eq:gain_approx}
k_{i}^{ f \text{ or }b} &=k^{ f \text{ or }b} (x)|_{x =\frac{N-i}{N}}, & b_{i}^{ f \text{ or }b} &=b^{ f \text{ or }b} (x)|_{x = \frac{N-i}{N}}, & m_i=m(x)|_{x = \frac{N-i}{N}}.
\end{align}
In addition, we define functions $k^{f+b}(x)$, $k^{f-b}(x)$, $b^{f+b}(x)$, $b^{f-b}(x):[0,1]^D\to \R$ as
\begin{align*}
  k^{f+b}(x) & \eqdef k^{f}(x) + k^{b}(x),  &  k^{f-b}(x) & \eqdef k^{f}(x) - k^{b}(x),\notag \\
  b^{f+b}(x) & \eqdef b^{f}(x)+ b^{b}(x),  &  b^{f-b}(x) & \eqdef b^{f}(x)- b^{b}(x).
\end{align*}
Due to~\eqref{eq:gain_approx}, these satisfy
\begin{align*}
 k_i^{f+b} &=k^{f+b}(x)|_{x =(N-i)/N}, &
k_i^{f-b} &=k^{f-b}(x)|_{x = (N-i)/N} \\
b_i^{f+b} &=b^{f+b}(x)|_{x = (N-i)/N}, &
b_i^{f-b} &=b^{f-b}(x)|_{x = (N-i)/N}.
\end{align*}
To obtain a PDE model from~\eqref{eq:prePDExx}, we first rewrite it as
\begin{align}\label{eq:prePDE2}
	m_i\ddot{\tilde{p}}_{i} =& \frac{k_i^{f-b}}{N}\frac{(\tilde{p}_{i-1}-\tilde{p}_{i+1})}{2(1/N)} +\frac{k_i^{f+b}}{2N^2}\frac{(\tilde{p}_{i-1}-2\tilde{p}_{i}+\tilde{p}_{i+1})}{1/N^2} \notag \\
	+& \frac{b_i^{f-b}}{N}\frac{(\dot{\tilde{p}}_{i-1}-\dot{\tilde{p}}_{i+1})}{2(1/N)} +\frac{b_i^{f+b}}{2N^2}\frac{(\dot{\tilde{p}}_{i-1}-2\dot{\tilde{p}}_{i}+\dot{\tilde{p}}_{i+1})}{1/N^2}.
\end{align}
Using the following finite difference approximations:
\begin{align*}
\Big[ \frac{\tilde{p}_{i-1}-\tilde{p}_{i+1}}{2(1/N)} \Big ] =\Big [\frac{\partial \tilde{p}(x,t)}{\partial x} \Big ]_{x = (N-i)/N},  \quad
\Big [ \frac{\tilde{p}_{i-1}-2\tilde{p}_{i}+\tilde{p}_{i+1}}{1/N^2} \Big ]
=\Big [\frac{\partial^2 \tilde{p}(x,t)}{\partial {x}^2} \Big ]_{x = (N-i)/N},\\
\Big[ \frac{\dot{\tilde{p}}_{i-1}-\dot{\tilde{p}}_{i+1}}{2(1/N)} \Big ] =\Big [\frac{\partial^2 \tilde{p}(x,t)}{\partial x \partial t} \Big ]_{x = (N-i)/N}, \quad
\Big [ \frac{\dot{\tilde{p}}_{i-1}-2\dot{\tilde{p}}_{i}+\dot{\tilde{p}}_{i+1}}{1/N^2} \Big ]
=\Big [\frac{\partial^3 \tilde{p}(x,t)}{\partial {x}^2 \partial t} \Big ]_{x = (N-i)/N}.
\end{align*}
For large $N$, Eq.~\eqref{eq:prePDE2} can be seen as a finite difference discretization of the following PDE:
\begin{align*}
	m(x) \frac{\partial^2 \tilde{p}(x,t)}{\partial t^2}   = \Big (\frac{k^{f-b}(x)}{N}\frac{\partial}{\partial x}+\frac{k^{f+b}(x)}{2{N}^2}\frac{\partial ^2}{\partial {x}^2}+ \frac{b^{f-b}(x)}{N}\frac{\partial^2}{\partial x \partial t}+\frac{b^{f+b}(x)}{2{N}^2}\frac{\partial ^3}{\partial {x}^2 \partial t} \Big ) \tilde{p}(x,t).
\end{align*}
The boundary conditions of the above PDE depend on the arrangement of reference agent in the redrawn graph of the network. For our case, the boundary condition is of Dirichlet type at $x=1$ where the reference agent is, and of Neumann type at $x=0$:
\begin{align*}
&\tilde{p}(1,t) = 0,& \frac{\partial \tilde{p}}{\partial x}(0,t) = 0.
\end{align*}

\section{Role of heterogeneity on stability margin}\label{sec:instability-analysis}
The starting point of our analysis is the investigation of the homogeneous and symmetric case: $m_i = m_0, k_i^{(\cdot)} = k_0,b_i^{(\cdot)} = b_0$ for some positive constants $m_0,k_0,b_0$, where $i \in \{1,\dots, N\}$. The  analysis leading to the proof of Theorem~\ref{thm:symmetric} is carried out using the PDE model derived in the previous section. In the homogeneous and symmetric control case, using the notation introduced earlier, we get 
\begin{align*}
	m(x)=m_0,\quad k^{f+b}(x) = 2k_0, \quad k^{f-b}(x) =0,\quad b^{f+b}(x) = 2b_0, \quad b^{f-b}(x) =0.
\end{align*}
The PDE~\eqref{eq:PDE_pre} simplifies to:
\begin{align}\label{eq:PDE_unpert}
	m_0\frac{\partial^2 \tilde{p}(x,t)}{\partial t^2} = \frac{k_0}{{N}^2}\frac{\partial ^2 \tilde{p}(x,t)}{\partial {x}^2}+\frac{b_0}{N^2}\frac{\partial ^3 \tilde{p}(x,t)}{\partial {x}^2 \partial t}.
\end{align}
This is wave equation with Kelvin-Voigt damping. Due to the linearity and homogeneity of the above PDE and boundary conditions, we are able to apply the method of separation of variables. We assume a solution of the form $\tilde{p}(x,t)=\sum_{\ell=1}^{\infty}\phi_\ell(x)h_{\ell}(t)$. Substituting it into PDE~\eqref{eq:PDE_unpert}, we obtain the following time-domain ODE
\begin{align}\label{eq:ODE_time}
	m_0\frac{d^2 h_\ell(t)}{d t^2}+\frac{b_0\lambda_\ell}{N^2}\frac{d h_\ell(t)}{d t} +\frac{k_0\lambda_\ell}{N^2} h_\ell(t)=0,
\end{align}
where $\lambda_\ell$ solves the following boundary value problem
\begin{align}\label{eq:ODE_space}
 \frac{d^2 \phi_\ell(x)}{d x^2}+\lambda_\ell  \phi_\ell(x)=0,
\end{align}
with the following boundary condition, which comes from~\eqref{eq:BC_pre}:
\begin{align}\label{eq:BC_SL}
	\frac{d \phi_\ell}{d x}(0) = 0,\quad \phi_\ell(1) =0. 
\end{align}
Following straightforward algebra, the eigenvalues and eigenfunction of the above boundary value problem is given by
 (see~\cite{haberman} for a BVP example)
\begin{align}\label{eq:lam}
  \lambda_\ell = \pi^2 \frac{(2\ell-1)^2}{4} , \quad \phi_\ell(x) = cos(\frac{2\ell - 1}{2}\pi x), \quad \ell=1,2,\cdots.
\end{align}
Take Laplace transform to both sides of the~\eqref{eq:ODE_time} with respect to the time variable $t$,  we obtain the characteristic equation of the PDE~\eqref{eq:PDE_unpert}:
\begin{align*}
m_0s^2 + \frac{b_0 \lambda_\ell}{N^2}s +\frac{k_0 \lambda_\ell}{N^2}   = 0.
\end{align*}
The eigenvalues of the PDE~\eqref{eq:PDE_unpert} are now given by
\begin{align}\label{eq:eigenvalue}
s_\ell^{\pm} =  - \frac{\lambda_\ell b_0}{2m_0N^2} \pm \frac{1}{2m_0N}\sqrt{\frac{\lambda_\ell^2b_0^2}{N^2} - 4\lambda_\ell m_0k_0}
\end{align}
For small $\ell$ and large $N$ so that $N>(2\ell-1)\pi b_0/(4\sqrt{m_0 k_0})$, the discriminant is negative, making the real part of the eigenvalues equal to $-\lambda_\ell b_0/(2m_0 N^2)$. The least stable eigenvalue, the one closest to the imaginary axis, is obtained with $\ell=1$:
\begin{align}\label{eq:s-min}
s_1^{\pm} = -\frac{\pi^2b_0}{8m_0}\frac{1}{N^2}+\Im \quad \Rightarrow \quad S \eqdef |Real(s_1^{\pm})| =\frac{\pi^2b_0}{8m_0N^2},
\end{align}
where $\Im$ is an imaginary number.
\smallskip

We are now ready to present the proof of Theorem~\ref{thm:symmetric}. 
\begin{proof-theorem}{\ref{thm:symmetric}}
Recall that in case of symmetric control we have
\begin{align*}
	\quad k_{i}^{f} =k_{i}^{b} , \quad b_{i}^{f} =b_{i}^{b}, \quad \forall i \in \{1,\cdots, N\}.
\end{align*}
In this case, using the notation introduced earlier, we have 
\begin{align*}
	k^{f-b}(x) =0, \quad b^{f-b}(x) =0, 
\end{align*}
The PDE~\eqref{eq:PDE_pre} is simplified to:
\begin{align}\label{eq:PDE_simplified}
	m(x)\frac{\partial^2 \tilde{p}(x,t)}{\partial t^2} = \frac{k^{f+b}(x)}{{2N}^2}\frac{\partial ^2 \tilde{p}(x,t)}{\partial {x}^2}+\frac{b^{f+b}(x) }{2N^2}\frac{\partial ^3 \tilde{p}(x,t)}{\partial {x}^2 \partial t},
\end{align}
The proof proceeds by a perturbation method. To be consistent with the bounds of the mass and control gains of each agent, let 
\begin{align*}
m(x)&=m_0+\delta \tilde{m}(x), \quad \tilde{m}(x) \in [-m_0,m_0] \\ k^{f+b}(x)&=2k_0+ \delta \tilde{k}(x),\quad \tilde{k}(x) \in [-2k_0,2k_0]\\ b^{f+b}(x)&=2b_0+\delta \tilde{b}(x),\quad \tilde{b}(x) \in [-2b_0,2b_0].
\end{align*}
where $\delta$ is a small positive number, denoting the amount of heterogeneity and $\tilde{m}(x), \tilde{k}(x), \tilde{b}(x)$ are the perturbation profiles. Take Laplace transform to both sides of  PDE~\eqref{eq:PDE_simplified} with respect to $t$, we have
\begin{align}\label{eq:lap_per}
	m(x)s^2  \eta = \frac{k^{f+b}(x)}{{2N}^2}\frac{\partial ^2 \eta}{\partial {x}^2}+\frac{b^{f+b}(x) }{2N^2}s\frac{\partial ^2 \eta}{\partial {x}^2 },
\end{align}
Let the perturbed eigenvalue be $s=s_{\ell}= s^{(0)}_{\ell} + \delta s^{(\delta)}_{\ell},$ the Laplace transform of $\tilde{p}(x,t)$ be $\eta = \eta^{(0)} + \delta \eta^{(\delta)}$, where $s^{(0)}_{\ell}$ and $\eta^{(0)}$ correspond to the unperturbed PDE~\eqref{eq:PDE_unpert}, i.e.
\begin{align}\label{eq:lap_unper}
	m_0(s^{(0)})^2  \eta^{(0)} = \frac{k_0}{N^2}\frac{\partial ^2 \eta^{(0)}}{\partial {x}^2}+\frac{b_0 }{N^2}s^{(0)}\frac{\partial ^2 \eta^{(0)}}{\partial {x}^2 }.
\end{align}
Eq.~\eqref{eq:eigenvalue} provides the formula for $s^{(0)}_{\ell}$ (actually, $s_{\ell}^\pm$), and $\eta^{(0)}$ is the solution to above equation, which is given by $\eta^{(0)}=\sum_{\ell=1}^\infty \eta^{(0)}_{\ell}=\sum_{\ell=1}^\infty \phi_\ell(x) H_\ell(s)$, where $H_{\ell}(s)$ is the Laplace transform of $h(t)$ given in~\eqref{eq:ODE_time}.
Plugging the expressions for $s_{\ell}$ and $\eta$ into~\eqref{eq:lap_per}, and doing an $O(1)$ balance leads to the eigenvalue equation for the unperturbed PDE, which is exactly Eq.~\eqref{eq:lap_unper}:
\begin{align*}
 \scr{P}\eta^{(0)}= 0, \text { where }  \scr{P} :=  \left( m_0(s^{(0)}_{\ell})^2 - \frac{b_0 s^{(0)}_{\ell}+ k_0}{N^2} \frac{\partial^2}{\partial x^2}  \right )
 \end{align*}
Next we do an $O(\delta)$ balance, which leads to:
\begin{align*}
	\scr{P}\eta^{(\delta)} = \Big(-2m_0 s_{\ell}^{(0)} s^{(\delta)}_{\ell}\eta^{(0)} -\tilde{m}(x){(s_{\ell}^{(0)})}^2 \eta^{(0)} + {\frac{\tilde{k}(x)}{2N^2} \frac{\partial^2 \eta^{(0)}}{\partial x^2}} + s_{\ell}^{(0)} {\frac{\tilde{b}(x)}{2N^2} \frac{\partial^2 \eta^{(0)} }{\partial x^2}}   +s^{(\delta)}_{\ell}  {\frac{b_0}{N^2} \frac{\partial^2 \eta^{(0)}}{\partial x^2}}   \Big)=: R
\end{align*}
For a solution $\eta^{(\delta)}$ to exist, $R$ must lie in the range space of the operator $\scr{P}$. Since $\scr{P}$ is self-adjoint, its range space is orthogonal to its null space. Thus, we have,
\begin{align}\label{eq:inner-product}
	< R , \eta_{\ell}^{(0)}> = 0
\end{align}
where $\phi_{\ell}$ is also the $\ell^{\text{th}}$ basis vector of the null space of operator $\scr{P}$. We now have the following equation:
\begin{align*}
\int_{0}^{1} \Big(-2m_0 s_{\ell}^{(0)} s^{(\delta)}_{\ell}\eta^{(0)} -\tilde{m}(x){(s_{\ell}^{(0)})}^2 \eta^{(0)}+ {\frac{\tilde{k}(x)}{2N^2} \frac{\partial^2 \eta^{(0)}}{\partial x^2}}  +s_{\ell}^{(0)} {\frac{\tilde{b}(x)}{2N^2} \frac{\partial^2 \eta^{(0)}}{\partial x^2}}   +s^{(\delta)}_{\ell}  {\frac{b_0}{N^2} \frac{\partial^2 \eta^{(0)}}{\partial x^2}}  \Big)\eta_{\ell}^{(0)} dx=0.
\end{align*}
Following straightforward manipulations, we got:
\begin{align}\label{eq:eig_pert2}
	s^{(\delta)}_{\ell} =\frac{b_0 \lambda_{\ell}}{m_0^2N^2} \int_0^1 \tilde{m}(x)(\phi_{\ell}(x))^2 dx  - \frac{\lambda_{\ell}}{2m_0N^2} \int_0^1 \tilde{b}(x)(\phi_{\ell}(x))^2 dx + \Im,
\end{align}
where $\Im$ is an imaginary number when $N$ is large ($N>(2\ell-1)\pi b_0/(4\sqrt{m_0 k_0})$). Using this, and substituting the equation above into $s_{\ell}=s^{(0)}_{\ell}+\delta s^{(\delta)}_{\ell}+O(\delta^2)$, and setting $\ell=1$, we obtain the stability margin of the heterogeneous network:
\begin{align*}
 S = \frac{b_0 \pi^2}{8m_0N^2} -\delta \frac{b_0 \pi^2}{4m_0^2N^2} \int_0^1 \tilde{m}(x)\cos^2 \big( \frac{\pi }{2}x \big ) dx  + \delta \frac{\pi^2}{8m_0N^2} \int_0^1 \tilde{b}(x)\cos^2 \big( \frac{\pi }{2}x \big )  dx +O(\delta^2).
\end{align*}
Plugging the bounds $|\tilde{m}(x)| \leq m_0$ and $|\tilde{b}(x)| \leq 2b_0$ , we obtain the desired result. \frQED
\end{proof-theorem}

\subsection{Numerical comparison}
\begin{figure}
	\psfrag{Real}{Real}
	\psfrag{Imaginary}{Imaginary}
	\psfrag{SSM}{SSM}
	\psfrag{PDE}{PDE}
\begin{center}
{\includegraphics[scale = 0.4]{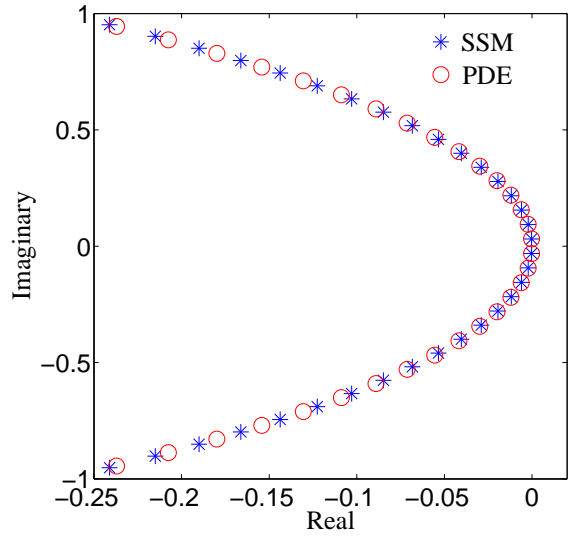}}
\end{center}
\caption{Numerical comparison of closed-loop eigenvalues of the state-space model (SSM) ~\eqref{eq:closedloop-wholeplatoon} and PDE model~\eqref{eq:PDE_simplified} with symmetric control. Eigenvalues shown are for a network of $50$ agents, and the mass and control gains profile are given in~\eqref{eq:gains}. Only some eigenvalues close to the imaginary axis are compared in the figure.}\label{fig:Eig_comparison}
\end{figure}

\begin{figure}
	\psfrag{N}{$\quad N$}
	\psfrag{S}{$S$}
	  \psfrag{SSM}{SSM}
	  \psfrag{PDE}{PDE}
	  \psfrag{Lower bound in (4)}{Lower bound in~\eqref{eq:less_stable_eigenvalue}}
	   \psfrag{Upper Bound in (4)}{Upper bound in~\eqref{eq:less_stable_eigenvalue}}
\begin{center}
{\includegraphics[scale = 0.4]{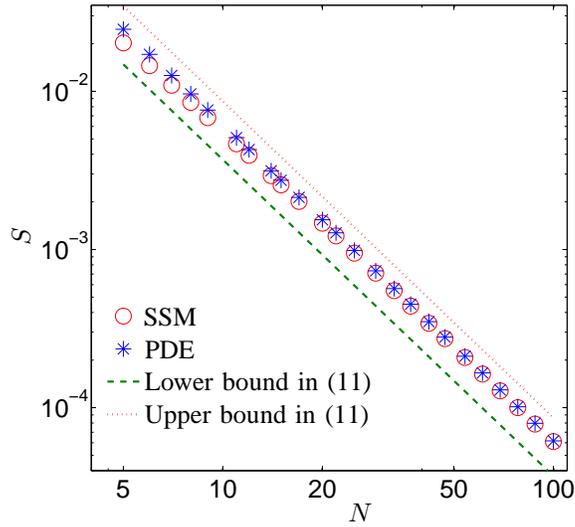}}
\end{center}
\caption{The stability margin of the heterogeneous formation with symmetric control as a function of number of agents: the legends of SSM, PDE and lower bound, upper bound stand for the stability margin computed from the  state space model, from the PDE model, and the asymptotic lower and upper bounds~\eqref{eq:less_stable_eigenvalue} in Theorem~\ref{thm:symmetric}. The mass and control gains profile are given in~\eqref{eq:gains}.
}\label{fig:Eig-trend}
\end{figure}

We now present numerical computations that corroborates the PDE-based analysis.  We consider the following mass and control gain profile:
\begin{align}\label{eq:gains}
	k_i^f = k^b_i &= 1+ 0.2\sin(2\pi (N-i)/N),\notag\\ b^f_i=b^b_i &= 0.5+0.1 \sin(2\pi (N-i)/N), \notag \\  m_i&=1+0.2\sin(2\pi (N-i)/N).
\end{align}
In the associated PDE model~\eqref{eq:PDE_simplified}, this corresponds to $k^f(x)=k^b(x)=1+0.2\sin(2\pi x)$, $b^f(x)=b^b(x)=0.5+0.1\sin(2\pi x)$, $m(x)=1+0.2\sin(2\pi x)$. The eigenvalues of the PDE, that are computed numerically using a Galerkin method with Fourier basis, are compared with that of the state space model to check how well the PDE model captures the closed loop dynamics. 
Figure~\ref{fig:Eig_comparison} depicts the comparison of eigenvalues of the state-space model and the PDE model. It shows the eigenvalues of the state-space model is accurately approximated by the PDE model, especially the ones close to the imaginary axis. We see from Figure~\ref{fig:Eig-trend} that the closed-loop stability margin of the controlled formation is well captured by the PDE model. In addition, the plot corroborates the predicted bound~\eqref{eq:less_stable_eigenvalue}.

\section{Role of asymmetry on stability margin}\label{sec:mistuning}
In this paper, we consider two scenarios of asymmetric control, we will first present the results when there is asymmetry in the velocity feedback alone (Theorem~\ref{thm:mistuning}). The results when there is equal asymmetry in both position and velocity feedback will follow immediately (Theorem~\ref{thm:equal-asymmetry}).
\subsection{Asymmetric velocity feedback}\label{sec:asymm_vel}
With symmetric control, one obtains an $O(\frac{1}{N^2})$ scaling law for the stability margin
because the coefficient of the $\frac{\partial ^3}{\partial {x}^2 \partial t}$ term  in the PDE~\eqref{eq:PDE_simplified} is $O(\frac{1}{N^2})$ and the coefficient of the $\frac{\partial^2}{\partial x \partial t}$ term  is $0$. Any asymmetry between the forward and the backward velocity gains will lead
to non-zero $b^{f-b}(x)$ and a presence of $O(\frac{1}{N})$ term as coefficient
of $\frac{\partial^2}{\partial x \partial t}$. By a judicious choice of asymmetry, there is
thus a potential to improve the stability margin from $O(\frac{1}{N^2})$
to $O(\frac{1}{N})$. A poor choice of control asymmetry may lead to instability, as we'll show in the sequel.

We begin by considering the forward and backward feedback gain profiles
\begin{align}\label{eq:kf-kf-tilde}
	k^f(x)=k^b (x)=k_0, \quad
	b^f(x)=b_0+\varepsilon \tilde{b}^f(x),\quad b^b (x)=b_0+\varepsilon \tilde{b}^b(x),
\end{align}
where $\varepsilon>0$ is a small parameter signifying the percent of asymmetry and $\tilde{b}^f(x)$, $\tilde{b}^b(x)$ are functions defined over $[0,1]$ that capture velocity gain perturbation from the nominal value $b_0$. Define
\begin{align}\label{eq:ks-km-def}
\tilde{b}^s(x) & \eqdef \tilde{b}^f(x)+ \tilde{b}^b(x), &
\tilde{b}^m(x) & \eqdef \tilde{b}^f(x)- \tilde{b}^b(x).
\end{align}
Due to the definition of $k^{f+b}, k^{f-b}, b^{f+b}$ and $b^{f-b}$, we have
\begin{align*}
k^{f+b}(x)&=2k_0, & k^{f-b}(x) & =0,\notag \\
b^{f+b}(x)&=2b_0+\varepsilon \tilde{b}^s(x), & b^{f-b}(x) & =\varepsilon \tilde{b}^m(x).
\end{align*}
The PDE~\eqref{eq:PDE_pre} with homogeneous mass $m_0$ now becomes
\begin{align}\label{eq:mistuned_PDE_D}
	m_0 \frac{\partial^2 \tilde{p}(x,t)}{\partial t^2} =
{\Big (  \frac{k_0}{N^2} \frac{\partial^2}{\partial x^2}}+ \frac{b_0}{N^2} \frac{\partial^3}{\partial x^2 \partial t} \Big) \tilde{p}(x,t)+  \varepsilon  \Big ( \frac{\tilde{b}^{s}(x)}{2N^2} \frac{\partial^3}{\partial x^2 \partial t} +\frac{\tilde{b}^{m}(x)}{N}\frac{\partial^2}{\partial x \partial t} \Big )\tilde{p}(x,t).
\end{align}

We now study the problem of how does the choice of the perturbations $\tilde{b}^{s}(x)$ and $\tilde{b}^{m}(x)$ (within limits so that the gains $b^f(x)$ and $b^b(x)$ are within pre-specified bounds) affect the stability margin. An answer to this question also helps in designing beneficial perturbations to improve the stability margin. The following result is used in the subsequent analysis. 
\begin{theorem}\label{thm:theorem1}
	Consider the eigenvalue problem of the PDE~\eqref{eq:mistuned_PDE_D} with mixed Dirichlet and Neumann boundary condition~\eqref{eq:BC_pre}. The least stable eigenvalue is given by the following formula that is valid for $\varepsilon \ll 1$ and large $N$:
\begin{align}\label{eq:less_stable_eig_mistuning_D}
   s_1  =s_1^{(0)}  -   \varepsilon \frac{\pi}{4m_0N} \int_{0}^{1} \tilde{b}^{m}(x)\sin \big(\pi x \big ) \ dx -  \varepsilon \frac{\pi^2}{8m_0N^2} \int_{0}^{1} \tilde{b}^{s}(x)\cos^2 \big(\frac{\pi}{2} x \big ) \ dx + O(\varepsilon^2)+\Im
\end{align}
where $s_1^{(0)}$ is the least stable eigenvalue of the unperturbed PDE~\eqref{eq:PDE_unpert} with the same boundary conditions and $\Im$ is an imaginary number when $N$ is large ($N>\pi b_0/(4\sqrt{m_0 k_0})$). \frqed
\end{theorem}

The proof of Theorem~\ref{thm:theorem1} is given in the Appendix. Now we are ready to prove Theorem~\ref{thm:mistuning}.
\begin{proof-theorem}{\ref{thm:mistuning}}
It follows from Theorem~\ref{thm:theorem1} that to minimize the least stable eigenvalue, one needs to choose only $\tilde{b}^m(x)$ carefully. The reason is the second term involving $\tilde{b}^s(x)$ has the $O(1/N^2)$ trend. Therefore, we choose
\begin{align*}
\tilde{b}^s(x) & \equiv 0.
\end{align*}
This means that the perturbations to the ``front'' and ``back'' velocity gains satisfy: 
\begin{align*}
\tilde{b}^f(x)=-\tilde{b}^b(x) \quad \quad \Leftrightarrow \quad \quad \tilde{b}^{m}(x)=2\tilde{b}^f(x).
\end{align*}
The most beneficial gains can now be readily obtained from Theorem~\ref{thm:theorem1}. To minimize the least stable eigenvalue with $\tilde{b}^s(x) \equiv 0$, we should choose $\tilde{b}^m(x)$ to make the integral $\int_{0}^{1} \tilde{b}^m(x)\sin(\pi x) dx$  as large as possible, which is achieved by setting $\tilde{b}^m(x)$ to be the largest possible value everywhere in the interval $[0, 1]$. The constraint $|b_i^{(\cdot)} - b_0|/b_0 \leq \varepsilon$ translates to $b_0(1-\varepsilon)  \leq b^{(\cdot)}(x) \leq b_0(1+\varepsilon)$, which means $\|\tilde{b}^f\|_\infty \leq b_0$ and $\|\tilde{b}^b\|_\infty \leq b_0$.  With the choice of $\tilde{b}^s$ made above, we therefore have the constraint $\|\tilde{b}^m\| \leq 2b_0$. The solution to the optimization problem is therefore obtained by choosing $\tilde{b}^m(x) = 2b_0$ $\forall x \in [0,1]$. This gives us the optimal gains
\begin{align*}
                    \tilde{b}^f(x)  =b_0, \quad \tilde{b}^b(x) =-b_0, \quad  
\Rightarrow \quad b^f(x)  = b_0(1+\varepsilon), \quad b^b(x)  = b_0(1-\varepsilon).
\end{align*}
The least stable eigenvalue is obtained from Theorem~\eqref{thm:theorem1}:
\begin{align*}
  s_1^+ = s^{(0)} - \frac{\varepsilon b_0}{m_0N} - 0 + O(\varepsilon^2)+ \Im.
\end{align*}
Since $s^{(0)}$ is the least stable eigenvalue for the symmetric PDE, we know from Theorem~\ref{thm:symmetric} that $s^{(0)} = O(1/N^2)$. Therefore, it follows from the equation above that the stability margin is $S = Re(s_1^+) = \frac{\varepsilon b_0}{m_0N} + O(\frac{1}{N^2})$. This proves the first statement of the theorem.

To prove the second statement, the control gain design $b_{i}^{f}=(1-\varepsilon)b_0$ and $b_{i}^{b}=(1+\varepsilon)b_0$ becomes $b^{f}(x)=(1-\varepsilon)b_0$ and $b^{b}(x)=(1+\varepsilon)b_0$. With this choice, it follows from Theorem~\eqref{thm:theorem1} that 
\begin{align*}
  s_1^+ = s^{(0)} + \frac{\varepsilon b_0}{m_0N} - 0 +O(\varepsilon^2)+ \Im.
\end{align*}
Since $s^{(0)} = O(1/N^2)$, the second term, which is $O(1/N)$, will dominate for large $N$. Since this term is positive, the second statement is proved. \frQED
\end{proof-theorem}

\subsection{Asymmetric position and velocity feedback with equal amount of asymmetry}\label{sec:asymm_pos_vel}
When there is equal asymmetry in the position and velocity feedback, we consider the following homogeneous and asymmetric control gains:
\begin{align}\label{eq:gains_equal}
k^f(x)&=(1+\varepsilon)k_0, & k^b(x) &=(1-\varepsilon)k_0,\notag \\
b^f(x)&=(1+\varepsilon)b_0, & b^b(x) & =(1-\varepsilon)b_0, 
\end{align}
where $\varepsilon$ is the amount of asymmetry satisfying $\varepsilon \in (0,1)$. 
\begin{proof-theorem}{\ref{thm:equal-asymmetry}}
The PDE model with the control gains specified in~\eqref{eq:gains_equal}  becomes
\begin{align}\label{eq:PDE_equal}
	m_0 \frac{\partial^2 \tilde{p}(x,t) }{\partial t^2}   = \frac{2\varepsilon k_0}{N}\frac{\partial \tilde{p}(x,t)}{\partial x}+\frac{k_0}{{N}^2}\frac{\partial ^2 \tilde{p}(x,t)}{\partial {x}^2}+ \frac{2\varepsilon b_0}{N}\frac{\partial^2 \tilde{p}(x,t)}{\partial x \partial t}+\frac{b_0}{{N}^2}\frac{\partial ^3 \tilde{p}(x,t)}{\partial {x}^2 \partial t} ,
\end{align}

By the method of separation of variables, we assume a solution of the form $\tilde{p}(x,t)=\sum_{\ell=1}^{\infty}\phi_\ell(x)h_{\ell}(t)$. Substituting it into PDE~\eqref{eq:PDE_equal}, we obtain the following time-domain ODE
\begin{align}\label{eq:ODE_time1}
	\frac{d^2 h_\ell(t)}{d t^2}+b_0\frac{d h_\ell(t)}{d t} +k_0\lambda_\ell h_\ell(t)=0,
\end{align}
where $\lambda_\ell$ solves the following boundary value problem
\begin{align}\label{eq:ODE_space}
  \L \phi_\ell(x)=0, \quad \L\eqdef \frac{d^2 }{d x^2}+2\varepsilon N\frac{d }{d x}+\lambda_\ell N^2,
\end{align}
with the following boundary condition, which comes from~\eqref{eq:BC_pre}:
\begin{align}\label{eq:BC_SL}
	\frac{d \phi_\ell}{d x}(0) = 0,\quad \phi_\ell(1) =0. 
\end{align}

Taking Laplace transform of both sides of~\eqref{eq:ODE_time1} with respect to the time variable $t$, we have the following characteristic equation for the PDE model
\begin{align}\label{eq:3}
 s^2 + b_0\lambda_\ell s + k_0\lambda_\ell = 0.
\end{align}
We now solve the boundary value problem~\eqref{eq:ODE_space}-\eqref{eq:BC_SL}. We first multiply both sides of~\eqref{eq:ODE_space} by $e^{2\varepsilon N x}N^2$, we obtain the standard Sturm-Liouville eigenvalue problem
\begin{align}\label{eq:SLeigen}
	\frac{d }{d x}\Big(e^{2\varepsilon N x}\frac{d \phi_\ell(x)}{d x}\Big)+\lambda^{(\varepsilon)}_\ell N^2e^{2\varepsilon N x}\phi_\ell(x)=0.
\end{align}
According to Sturm-Liouville Theory, all the eigenvalues are real and have the following ordering $\lambda_1<\lambda_2<\cdots$, see~\cite{haberman}. To solve the boundary value problem~\eqref{eq:ODE_space}-\eqref{eq:BC_SL}, we assume solution of the form, $\phi_\ell(x)=e^{rx}$, then we obtain the following equation
\begin{align}\label{eq:chara_1}
	r^2+2\varepsilon N r+\lambda_\ell N^2=0, \quad
	\Rightarrow  \quad  r=-\varepsilon N\pm N\sqrt{\varepsilon^2-\lambda_\ell}.
\end{align}
Depending on the discriminant in the above equation, there are three cases to analyze:
\begin{enumerate}
\item $\lambda_\ell < \varepsilon^2$,  the eigenfunction has the following form $\phi_\ell(x)=c_1 e^{(-\varepsilon N+ N\sqrt{\varepsilon^2-\lambda_\ell})x}+c_2 e^{(-\varepsilon N- N\sqrt{\varepsilon^2-\lambda_\ell})x}$,
	where $c_1, c_2$ are some constants. Applying the boundary condition~\eqref{eq:BC_SL}, it's straightforward to see that, for non-trivial eigenfunctions $\phi_\ell(x)$ to exit,  the following equation must be satisfied $(\varepsilon N- N\sqrt{\varepsilon^2-\lambda_\ell})/(\varepsilon N+ N\sqrt{\varepsilon^2-\lambda_\ell})=e^{2N\sqrt{\varepsilon^2-\lambda_\ell}}$. For positive $\varepsilon$, this leads to a contradiction, so there is no eigenvalue for this case.
\item $\lambda_\ell =\varepsilon^2$,  the eigenfunction $\phi_\ell(x)$ has the following form
		\begin{align*}
			\phi_\ell(x)=c_1 e^{-\varepsilon N x}+c_2 x e^{-\varepsilon N x}.
		\end{align*}
 Again, applying the boundary condition~\eqref{eq:BC_SL}, for non-trivial eigenfunctions $\phi_\ell(x)$ to exit,  we have the following $\varepsilon N=-1$, which implies there is no eigenvalue for this case either.
\item $\lambda_\ell >\varepsilon^2$,  the eigenfunction has the following form $\phi_\ell(x)= e^{-\varepsilon N x} (c_1 \cos(N\sqrt{\lambda_\ell-\varepsilon^2} x)+c_2 \sin(N\sqrt{\lambda_\ell-\varepsilon^2} x))$.  Applying the boundary condition~\eqref{eq:BC_SL}, for non-trivial eigenfunctions $\phi_\ell(x)$ to exit, the eigenvalues $\lambda_\ell$ must satisfy $\lambda_\ell= \varepsilon^2+\frac{a_\ell^2}{N^2}$ where $a_\ell$ solves the transcendental equation $-a_\ell/(\varepsilon N)=\tan(a_\ell)$. A graphical representation of the functions $\tan x$ and $-x/\varepsilon N$ with respect to $x$ shows that $a_\ell \in (\frac{(2\ell-1)\pi}{2}, \ell \pi)$. 
\end{enumerate} 

From case 3), we see that $a_1 \in (\pi/2, \pi)$, and $\lambda_1 \to \varepsilon^2$ from above as $N \to \infty$, i.e. $\inf_N \lambda_1= \varepsilon^2$. For each $\ell \in \{1,2,\cdots \}$, the two roots of the characteristic equations~\eqref{eq:3} are given by
	\begin{align}\label{eq:eigenvalue_least}
		s_{\ell}^{\pm} = \frac{- b_0\lambda_\ell \pm \sqrt{b_0^2\lambda_\ell^2-4k_0\lambda_\ell}}{2}.
 \end{align}
Depending on the discriminant in~\eqref{eq:eigenvalue_least}, there are two cases to analyze:
\begin{enumerate}
\item If $ \lambda_1 \geq 4k_0/b_0^2$, then the discriminant in~\eqref{eq:eigenvalue_least} for each $\ell$ is non-negative, the \emph{less stable} eigenvalue can be written as
\begin{align*}
	s_\ell^+=-\frac{\lambda_{\ell} b_0-\sqrt{(\lambda_{\ell}b_0)^2-4\lambda_{\ell}k_0}}{2}= -\frac{2k_0}{b_0+\sqrt{b_0^2-4k_0/\lambda_\ell}}.
\end{align*}
The least stable eigenvalue is achieved by setting $\lambda_\ell=\lambda_\infty$. Since $\lambda_\ell \to \infty$ as $\ell \to \infty$, we have the stability margin
\begin{align*}
	S=|Re(s_{1}^+)|\geq \frac{2k_0}{b_0+\sqrt{b_0^2-0}}=\frac{k_0}{b_0}.
\end{align*}
\item Otherwise, the discriminant in~\eqref{eq:eigenvalue_least} is indeterministic, i.e. it's negative for small $\ell$ and positive for large $\ell$ is non-positive. For those $\ell$'s which make the  discriminant negative, the least stable eigenvalue among them is given by
\begin{align*}
	s_1^{\pm}=-\frac{\lambda_1 b_0}{2}+\Im.
\end{align*}
where $\Im$ is an imaginary number. For those $\ell$'s which make the discriminant non-positive, we have from Case 1) that the least stable eigenvalue among them is given by
\begin{align*}
s_1^+=-\frac{2k_0}{b_0+\sqrt{b_0^2-4k_0/\lambda_\infty}}
\end{align*}
The stability margin is given by taking the minimum of absolute value of the real part of the above two eigenvalues,\begin{align*}
S\geq	\min \Big \{ \frac{b_0 \lambda_1}{2}, \frac{k_0}{b_0} \Big\}.
\end{align*}
\end{enumerate} 
Combining the above two cases, and using the fact that $\lambda_1 \geq \varepsilon^2$, we obtain that the stability margin can be bounded below as follows
\begin{align*}
S\geq	\min \Big \{ \frac{b_0 \varepsilon^2 }{2}, \frac{k_0}{b_0} \Big\}.
\end{align*}
This concludes the proof. \frQED

\end{proof-theorem}

\subsection{Numerical comparison of stability margin}

\begin{figure}[t]
	\psfrag{N}{$\quad N$}
	\psfrag{S}{$S$}
        \psfrag{data1}{\small Symmetric (SSM)}
 	\psfrag{data2}{\small Symmetric (PDE)}
        \psfrag{data3}{\small Asymmetric velocity (SSM)}
 	\psfrag{data4}{\small Asymmetric velocity (PDE)}
	\psfrag{data5}{\small Asymmetric position and velocity (SSM)}
 	\psfrag{data6}{\small Asymmetric position and velocity (PDE)}
	\psfrag{data7}{\small Theorem~\ref{thm:mistuning}}
 	\psfrag{data8}{\small Theorem~\ref{thm:equal-asymmetry}}
\begin{center}
{\includegraphics[scale = 0.5]{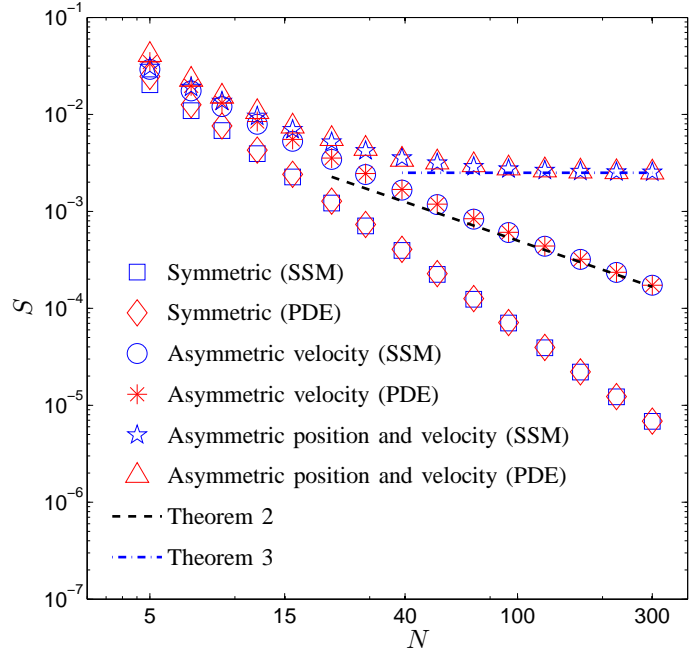}}
\end{center}
\caption{Stability margin improvement by asymmetric control. The mass of each agent used is $m_0=1$. The nominal control gains are $k_0=1$, $b_0=0.5$. The asymmetric control gains used are the ones given  in Theorem~\ref{thm:mistuning} and  Theorem~\ref{thm:equal-asymmetry} respectively, and the amount of asymmetry is $\varepsilon=0.1$. The legends ``SSM'' and ``PDE'' stand for the stability margin computed from the state-space model and the PDE model, respectively. }\label{fig:mistuned-lseig-compare}
\end{figure}

Figure~\ref{fig:mistuned-lseig-compare} depicts the numerically obtained  stability margins for both the PDE and state-space models (SSM) with symmetric and  asymmetric control gains. The figure shows that 1) the stability margin of the PDE model  matches that of the state-space model accurately, even for small values of $N$; 2) the stability margin with asymmetric velocity feedback shows large improvement over the symmetric case even though the velocity gains differ from their nominal values only by $\pm 10\%$. The improvement is particularly noticeable for large values of $N$, while being significant even for small values of $N$; 3) With equal amount of asymmetry in both the position and velocity feedback, the stability margin can be uniformly bounded away from $0$, which eliminates the degradation of stability margin with increasing $N$; 4)  the asymptotic  formulae given in Theorem~\ref{thm:mistuning} and Theorem~\ref{thm:equal-asymmetry} are quire accurate. 

\begin{figure}[t]
	\psfrag{N}{$N$}
	\psfrag{S}{$Re(s^+_1)$}
	  \psfrag{SSM}{\small Poor asymmetric velocity  (SSM) }
	   \psfrag{PDE}{\small Poor asymmetric velocity  (PDE) }
	   \psfrag{eq}{\small Theorem~\ref{thm:mistuning}}
\begin{center}
{\includegraphics[scale = 0.4]{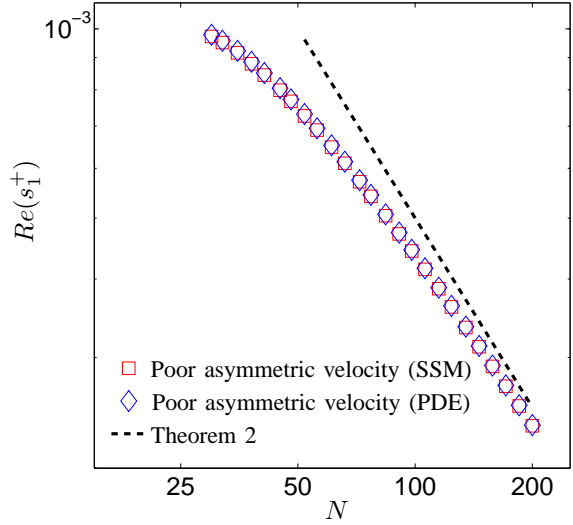}}
\end{center}
\caption{The real part of the most unstable eigenvalues with poor asymmetry. The mass of each agent is $m_0=1$. The nominal control gains are $k_0=1$, $b_0=0.5$, and the control gains used are the ones given by~\eqref{eq:bad-mistuned-gains} in Theorem~\ref{thm:mistuning} with $\varepsilon=0.1$. }\label{fig:poor_asymmetry}
\end{figure}

Numerical validation that poor choice of asymmetry in control gains can lead to instability is shown in Figure~\ref{fig:poor_asymmetry}. Note that the real part of these eigenvalues are positive and Eq.~\eqref{eq:bad-mistuned-gains}  also makes an accurate prediction.

\subsection{Sensitivity to external disturbances}
When external disturbances are present, we model the dynamics of vehicle $i$ by $\ddot{p}_i=\ddot{\tilde{p}}_i = u_i +  w_i$, where $w_i$ is the external disturbance acting on the vehicle and the mass of each vehicle is assumed to be $m_0=1$.  Each component of the disturbance is assumed to be independent. Then the closed-loop dynamics of the formation is given by
\begin{align}\label{eq:error}
\dot{ \mbf \psi} & = {\mbf A}{\mbf \psi} + B w, \quad B= I_N \otimes \matt{0 \\ 1},
\end{align}
where $\psi \eqdef[\tilde{p}_{1},\dot{\tilde p}_{1}, \tilde{p}_{2},\dot{\tilde p}_{2},\cdots,\tilde{p}_{N}, \dot{\tilde p}_{N}]^{T}$ is the state vector, $w  \eqdef [w_1,w_2,\dots,w_N]^T$ is the vector of disturbances and $I_N$ is the $N \times N$ identity matrix. We consider the vector of errors $e \eqdef
 [\tilde{p}_1,\dots,\tilde{p}_N]^T$ as the outputs:
 \begin{align*}
e & = C \mbf{\psi} ,\quad C= I_N \otimes [0 ,\ 1].
 \end{align*}
The $H_\infty$ norm of the transfer function $G_{we}$ from the disturbance $w$ to the errors $e$ is a measure of the closed-loop's sensitivity to external disturbance. 

\begin{figure}[t]
	\psfrag{N}{$\quad N$}
	\psfrag{H8norm}{$\ \ || G_{we} ||_{\infty}$}
	\psfrag{rho=0}{Symmetric control}
	\psfrag{rho=0.1}{Asymmetric position \& velocity}
	\psfrag{epsilon=0.1}{Asymmetric velocity}
	\psfrag{150}{}
\begin{center}
{\includegraphics[scale = 0.4]{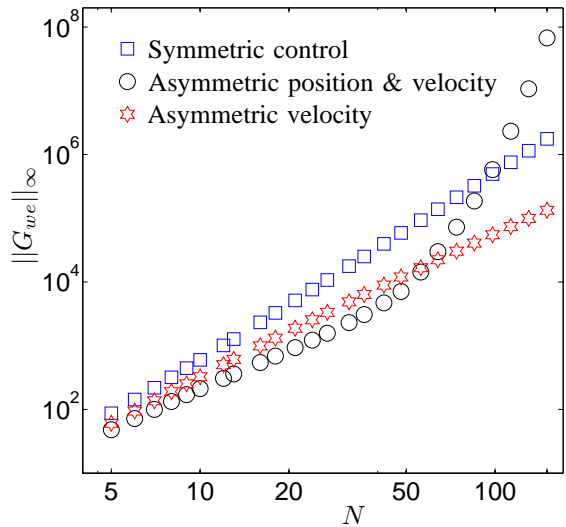}}
\end{center}
\caption{Numerical comparison of $H_\infty$ norm of the transfer function $G_{we}$ from disturbance $w$ to spacing error $e$. The mass of each agent is assumed homogeneous and is given by $m_0=1$. The nominal control gains used are $k_0=1, b_0=0.5$. The asymmetric control gains are given in Theorem~\ref{thm:mistuning} (asymmetric velocity) and Theorem~\ref{thm:equal-asymmetry} (asymmetric position and velocity) respectively. The amount of asymmetry used is $\varepsilon=0.1$.  Norms are computed using the Control Systems Toolbox in MATLAB$^\text{\copyright}$.}\label{fig:all_norm}
\end{figure}

Figure~\ref{fig:all_norm} depicts the $H_\infty$ norm of $G_{we}$ as a function of $N$.  It's surprising to see that with equal asymmetry in position and velocity feedback, the $H_\infty$ norm of $G_{we}$ blows up for large $N$. However, for asymmetric velocity feedback control , the norm is improved comparing to the symmetric control case. Therefore, in the sense of sensitivity to external disturbances, the asymmetric velocity feedback control exceeds symmetric control, which in turn exceeds asymmetric position and velocity feedback control with equal asymmetry. This results implies that asymmetric velocity feedback is the best choice for large stability margin and better sensitivity to disturbances. 

\begin{remark}\label{rem:veerman}
This result with equal asymmetry in the position and velocity feedback  is similar to Veerman's result (see~\cite{veerman_stability}), in which he consider the amplitude of the last agent in the network when only the reference agent experiences a harmonic disturbance. He concludes that this disturbance is amplified exponentially in $N$ as it propagates through the network when there is equal asymmetry in both position and velocity feedback and the growth is linear in $N$ for symmetric control. Our result is a complementary result, we show that when there is only asymmetry in velocity feedback, the disturbance amplification factor can be decreased, compared to symmetric control, which is  superior to asymmetric position and velocity feedback control. Analysis of these trends is beyond the scope of this work, and will be undertaken in future work.
\end{remark}

\section{Summary}\label{sec:conc}
We studied the role of heterogeneity and control asymmetry on the stability margin of a large 1D network of double-integrator agents. The control is in a distributed sense that the control signal at every agent depends on the relative position and velocity measurements from its two nearest neighbors (one one either side). 
It is shown that heterogeneity does not effect how the stability margin scales with $N$, the number of agents, whereas asymmetry plays a significant role. As long as control is symmetric, meaning information on relative position and velocity from both neighbors are weighed equally, agent-to-agent heterogeneity does not change the $O(1/N^2)$ scaling of stability margin. If front-back asymmetry is introduced in the control gains, even by an arbitrarily small amount, the stability margin can be improved to $O(1/N)$ with asymmetric velocity feedback. The stability margin can be even improved to $O(1)$ if there is equal amount of asymmetry in the position and velocity feedback. However, from the perspective of sensitivity to disturbances, numerical simulation shows that  asymmetric position and velocity feedback has much worse performance than asymmetric velocity feedback, and asymmetric velocity feedback exceeds symmetric control. Therefore, the asymmetric velocity feedback scheme provides a best way to achieve the goal of larger stability margin and better sensitivity to external disturbances. The scenarios with unequal asymmetry in position and velocity feedback and asymmetric position feedbacks are open problems.

\bibliographystyle{IEEEtran}
\bibliography{../../PBbib/vehicular_platoon,../../PBbib/sensnet_bib_dbase,../../PBbib/Barooah,../../PBbib/HH,../../PBbib/distributed_control}

\appendix

\begin{proof-theorem}{\ref{thm:theorem1}}
The proof proceeds by a perturbation method. Let the eigenvalues and Laplace transformation of $\tilde{p}(x,t)$ for the perturbed PDE~\eqref{eq:mistuned_PDE_D} be
$
s_{\ell}= s^{(0)}_{\ell} + \varepsilon s^{(\varepsilon)}_{\ell}, \eta = \eta^{(0)} + \varepsilon \eta^{(\varepsilon)}
$
respectively, where $s^{(0)}_{\ell}$ and $\eta^{(0)}$ are corresponding to the unperturbed PDE~\eqref{eq:PDE_unpert}. 
Taking a Laplace transform of PDE~\eqref{eq:mistuned_PDE_D}, plugging in the expressions for $s_{\ell}$ and $\eta$, and doing an $O(\varepsilon)$ balance, which leads to:
\begin{align*}
	\scr{P}\eta^{(\varepsilon)}= s_{\ell}^{(0)} {\frac{\tilde{b}^{m}(x)}{N}\frac{d \eta^{(0)}}{d x}}  + s_{\ell}^{(0)} {\frac{\tilde{b}^{s}(x)}{2N^2} \frac{d^2 \eta^{(0)}}{d x^2}}  -2 m_0s_{\ell}^{(0)} s^{(\varepsilon)}_{\ell}\eta^{(0)} +s^{(\varepsilon)}_{\ell}  \frac{b_0}{N^2} \frac{d^2 \eta^{(0)}}{d x^2}=: R
\end{align*}
For a solution $\eta^{(\varepsilon)}$ to exist, $R$ must lie in the range space of the self-adjoint operator $\scr{P}$. Thus, we have,
\begin{align*}
	< R , \eta^{(0)}_{\ell}> = 0
\end{align*}
We now have the following equation:
\begin{align*}
\int_{0}^{1}  \Big( s_{\ell}^{(0)} {\frac{\tilde{b}^{m}(x)}{N}\frac{d \eta^{(0)}}{d x}} + s_{\ell}^{(0)} {\frac{\tilde{b}^{s}(x)}{2N^2} \frac{d^2 \eta^{(0)}}{d x^2}}  -2 m_0s^{(0)}_{\ell} s^{(\varepsilon)}_{\ell} \eta^{(0)}+  s^{(\varepsilon)}_{\ell}  \frac{b_0}{N^2} \frac{d^2 \eta^{(0)}}{d x^2} \Big) \eta^{(0)}_{\ell} dx=0
\end{align*}
Following straightforward manipulations, we get:
\begin{align}\label{eq:eig_pert1}
	m_0( s^{(0)}_{\ell}+ \frac{b_0\lambda_{\ell}}{2m_0N^2} )s^{(\varepsilon)}_{\ell} =&-s^{(0)}_{\ell}\frac{(2\ell-1)\pi}{4 N} \int_{0}^{1} \tilde{b}^{m}(x)\sin \big((2\ell-1)\pi x \big )  \ dx \notag \\ &-s^{(0)}_{\ell}\frac{(2\ell-1)^2\pi^2}{8N^2}  \int_{0}^{1}  \tilde{b}^{s}(x)\cos^2 \big(\frac{(2\ell-1)\pi}{2} x \big ) \ dx.
\end{align}
Substituting the equation above into $s_{\ell}=s^{(0)}_{\ell}+\varepsilon s^{(\varepsilon)}_{\ell}$, and set $\ell=1$, we  complete the proof. \frQED
\end{proof-theorem}

\begin{proof-lemma}{\ref{lem:hetero_mass}}
With the  profiles  and control gains given in Lemma~\ref{lem:hetero_mass}, the PDE~\eqref{eq:PDE_pre} simplifies to:
\begin{align}\label{eq:PDE_unpert2}
	m(x)\frac{\partial^2 \tilde{p}(x,t)}{\partial t^2} = \frac{\rho b(x)}{{N}^2}\frac{\partial ^2 \tilde{p}(x,t)}{\partial {x}^2}+\frac{b(x)}{N^2}\frac{\partial ^3 \tilde{p}(x,t)}{\partial {x}^2 \partial t},
\end{align}
where $m_\mrm{min} \leq m(x) \leq m_\mrm{max}, b_\mrm{min} \leq b(x) \leq b_\mrm{max}$. Due to the linearity and homogeneity of the above PDE and boundary conditions, we are able to apply the method of separation of variables. We assume solution of the form $\tilde{p}(x,t)=\sum_{\ell=1}^\infty \phi_\ell(x)h_\ell(t)$. Substituting the solution into~\eqref{eq:PDE_unpert2} and dividing both sides by $\phi_\ell(x)h_\ell(s)$, we obtain:
\begin{align}\label{eq:separation}	
	\frac{h''_\ell(t)}{\frac{\rho}{N^2} h_\ell(t)+\frac{1}{N^2} h(t)}=\frac{\phi''_\ell(x)}{m(x)\phi_\ell(x)/b(x)}
\end{align}
Since each side of the above equation is independent from the other, so it's necessary for both sides equal to the same constant $-\lambda_{\ell}$. Then we have two separate equations:
\begin{align}
	 h''_\ell(t)+\frac{\lambda_{\ell}}{N^2}h'_\ell(t)+\frac{\rho\lambda_{\ell}}{N^2}h_\ell(t)=0, \\ \label{eq:charac}
	 \phi''(x)+\lambda_{\ell} \frac{m(x)}{b(x)}\phi(x)=0.
\end{align}
The spatial part solves the following regular Sturm-Liouville eigenvalue problem
\begin{align}\label{eq:xODE}
\phi''(x)+\lambda_{\ell} \frac{m(x)}{b(x)} \phi(x)=0, \notag\\
\frac{d \phi(0)}{dx}=\phi(1)=0.
\end{align}
The Rayleigh quotient is given by
\begin{align}\label{eq:RQ}
\lambda_{\ell}=\frac{\int_0^1 (d \phi(x) / dx)^2 dx}{\int_0^1 \phi^2(x) m(x)/b(x) dx}.
\end{align}
Since $m_\mrm{min} \leq m(x) \leq m_\mrm{max}, b_\mrm{min} \leq b(x) \leq b_\mrm{max}$, we have that $\frac{m_\mrm{min}}{b_\mrm{max}} \leq m(x)/b(x) \leq \frac{m_\mrm{max}}{b_\mrm{min}}$. Plugging the lower and upper bounds for $m(x)/b(s)$, we have the following relation:
\begin{align*}
\frac{b_\mrm{min}}{m_\mrm{max}} \frac{\int_0^1 (d \phi(x) / dx)^2 dx}{\int_0^1 \phi^2(x) dx} \leq \lambda_{\ell} \leq \frac{b_\mrm{max}}{m_\mrm{min}} \frac{\int_0^1 (d \phi(x) / dx)^2 dx}{\int_0^1 \phi^2(x) dx}
\end{align*}
Since we know the eigenvalue $\bar{\lambda}_{\ell}$ corresponding to Rayleigh quotient $\frac{\int_0^1 (d \phi(x) / dx)^2 dx}{\int_0^1 \phi^2(x) dx} $ is the eigenvalue obtained from~\eqref{eq:xODE} with $m(x)/b(x)=1$. And $\bar{\lambda}_{\ell}$ is given by
\begin{align}\label{eq:eigenvalue1}
\bar{\lambda}_{\ell}=\frac{(2\ell-1)^2\pi^2}{4}
\end{align}
where $\ell$ is the wave number, $\ell=1,2,\cdots$.

It is straight forward to see that the least eigenvalue $\bar{\lambda}_{\ell}$ is obtain by setting $\ell=1$, i.e. $\bar{\lambda}_1=\pi^2/4$. So we have the following bounds for the least eigenvalue of $\lambda_{\ell}$.
\begin{align}\label{eq:bound}
	\frac{b_\mrm{min}\pi^2}{4m_\mrm{max}} \leq \lambda_1 \leq \frac{b_\mrm{max}\pi^2}{4m_\mrm{min}} 
\end{align}
Take Laplace transform to both sides of~\eqref{eq:charac}, we obtain the following characteristic equation for the PDE model~\eqref{eq:PDE_unpert2}. 
\begin{align*}
	s^2 +\frac{\lambda_\ell}{N^2}s+\frac{\rho\lambda_\ell}{N^2}=0.
\end{align*}
Its eigenvalues turn out to be the roots of the above equation,
\begin{align}\label {eq:eigenvalue2}
 s_{\ell}^{\pm} \eqdef \frac{-\lambda_{\ell}/N^2 \pm\sqrt{ \lambda_{\ell}^2/N^4- 4\rho \lambda_{\ell}/N^2}}{2}.
\end{align}
We call $s_{\ell}^{\pm}$ the $\ell$-th pair of eigenvalues. The discriminant D in~\eqref{eq:eigenvalue2} is given by:
\begin{align*}
D \eqdef &\lambda_{\ell}^2/N^4- 4\rho \lambda_{\ell}/N^2.
\end{align*}
For large $N$ and small $\ell$, $D$ is negative. So both the eigenvalues in~\eqref{eq:eigenvalue2} are complex, then the stability margin is only determined by the real parts of $s_{\ell}^{\pm}$. 
It follows from~\eqref{eq:eigenvalue2} that the least stable eigenvalue (the ones closest to the imaginary axis) among them is the one that  is obtained by minimizing $\lambda_{\ell}$ over $\ell$.  Then, this minimum is achieved at $\ell=1$, and the real part  is obtained
 \begin{align*}
	 Real(s_1^{\pm})=-\frac{\lambda_1}{2N^2}.
 \end{align*}
 Following the definition of stability margin $S \eqdef |Real(s_1^{\pm})|$ as well as the bounds for $\lambda_1$ given by~\eqref{eq:bound}, we complete the proof. \frQED
\end{proof-lemma}

\end{document}